\documentclass[lettersize,journal]{IEEEtran}
\usepackage{amsmath,amsfonts}
\usepackage{algorithmic}
\usepackage{algorithm}
\usepackage{array}
\usepackage[caption=false,font=normalsize,labelfont=sf,textfont=sf]{subfig}
\usepackage{textcomp}
\usepackage{stfloats}
\usepackage{url}
\usepackage{verbatim}
\usepackage{graphicx}
\usepackage{multirow}
\usepackage{booktabs}
\usepackage{cite}
\usepackage{listings}
\title{\LARGE \bf
TokenPatronus: A Decentralized NFT Anti-theft Mechanism
}
\author{Zheng Cao, Yi Zhen, Gang Fan and Sheng Gao\\
 z.cao@zju.edu.cn, \{iamzhenyi, fan.gang.cn, jonathan.gonse\}@gmail.com
\thanks{*This work was not supported by any organization}
\thanks{Z. Cao is with the College of Computer Science and Technology, Zhejiang University, Hangzhou 310016, China}
}

\begin{document}

\maketitle
\thispagestyle{empty}
\pagestyle{empty}

\begin{abstract}
The emergence of metaverse brings tremendous evolution to Non-Fungible Tokens (NFTs), which could certify the ownership the unique digital asset in the cyber world. The NFT market has garnered unprecedented attention from investors and created billions of dollars in transaction volume. Meanwhile, securing NFT is still a challenging issue. Recently, numerous incidents of NFT theft have been reported, leading to incalculable losses for holders. We propose a decentralized NFT anti-theft mechanism called TokenPatronus, which supports the general ERC-721 standard and provide the holders with strong property protection. TokenPatronus contains pre-event protection, in-event interruption, and post-event replevin enhancements for the complete NFTs transactions stages. Four modules are designed to make up the decentralized anti-theft mechanism, including the decentralized access control (DAC), the decentralized risk management (DRM), the decentralized arbitration system (DAS) and the ERC-721G standard smart contract. TokenPatronus is performing on the Turtlecase NFT project of Ethereum and will support more blockchains in the future. 
\end{abstract}
\begin{IEEEkeywords}
	NFT, Blockchain, Security, Ethereum
\end{IEEEkeywords}

\section{Introduction}
The concept of the metaverse is mushrooming after COVID-19,  which is a hybrid of advanced technologies including Artificial Intelligence, Augmented Reality (AR), Block-chain, Cryptocurrencies, Internet of Things (IoT), and Non-Fungible Tokens (NFTs). However, considering the multi-layer nature of metaverse, the security issue is a crucial challenge in developing the related applications \cite{falchuk2018social,di2021metaverse}. NFT refers to a unit of data stored on the blockchain that can certify digital asset ownership. The content of NFT is not limited to image, music, code, domain, or other intellectual property, with the traceable, easily verifiable, and hard-to-tamper features \cite{kugler2021non}. Due to the above advantages, the NFT is considered the future of digital art, and it has generated over \$22 billion in sales \cite{kireyev2022nft}. However, compared with the cryptocurrencies like Bitcoin, NFT cannot be traded as the universal equivalent due to its unique feature. Thus, popular NFT marketplaces such as OpenSea and Nifty Gateway have appeared for simplifying the trading and popularizing the NFT culture \cite{gupta2022identifying}. Nowadays, in spite of the classic NFTs like cyberpunk collection, many traditional platforms have attempted to create their own NFTs, such as Disney, Nike and NBA. There's no doubt that the NFT market is in continuing growth and attracting more and more attentions \cite{das2021understanding,takahashi2021voting}.\\

Although non-fungible tokens are taking the digital world by storm with people making money and trading these digital assets, there are also plenty of scams to dupe people and businesses out of money. Since 2020, 166 thefts have been recorded and over \$86.6 million of tokens have been stolen\cite{stolennfts}. Recently, OpenSea has been reported security vulnerabilities that might have resulted in vital losses of the investors. This problem allows attackers to steal the customer's entire crypto wallet simply by sending a malicious NFT. The hackers may access the full crypto wallet of the victims by pop-up the fraudulent URLs and then and steal all their funds. Generally, hacker may attack the NFT system via the following methods\cite{fisher2019once,wang2021non,elzweig2022does}: 
\begin{itemize}
	\item Using sophisticated social engineering, such as phishing and fraudulent accounts pretending to be an administrator;
	\item Exploiting the vulnerabilities in the smart contract security, e.g., the cross-chain Defi site PolyNetwork was attacked via the flaw in their heterogeneous blockchains contract;
	\item Identity fraud, such as fake stores with the capability to mirror the logo and contents of the original NFT store. These counterfeits may be passed off as the real thing and sold, leading to a user buying a fake NFT, which leads to copyright issues since there can only be one NFT;
	\item Abuse of bot exploits like the Mee6 bot, which allows admins to automatically give and remove roles and send messages to the community. In some instances, the attackers even updated administrator settings to ban Discord moderators from interfering with the hackers’ operations.
\end{itemize}

However, current security solutions are still imperfect, and relying solely on users to raise security awareness is not enough to deal with the complex and changing hacking behavior. Moreover, security assistance features of trading markets such as Blue Label authentication provided by OpenSea, are limited to its centralized implementations. Therefore, we propose TokenPatronus as a decentralized solution to provide the trusted protection way to the NFT assets. Our solution mainly uses the Oracle as the decentralized means of the NFT security, including the decentralized access control (DAC), decentralized risk management (DRM) and decentralized arbitration system (DAS). Besides, we propose the ERC-721G smart contract standard to interact with the oracle, and eventually safeguard our NFT from theft. In addition, we have applied the implementation to the Turtlecase series NFTs \footnote{https://turtlecase.xyz/}. More details are described in the following sections.

\begin{figure*}[!hb]
\centering
\includegraphics[width=1\textwidth]{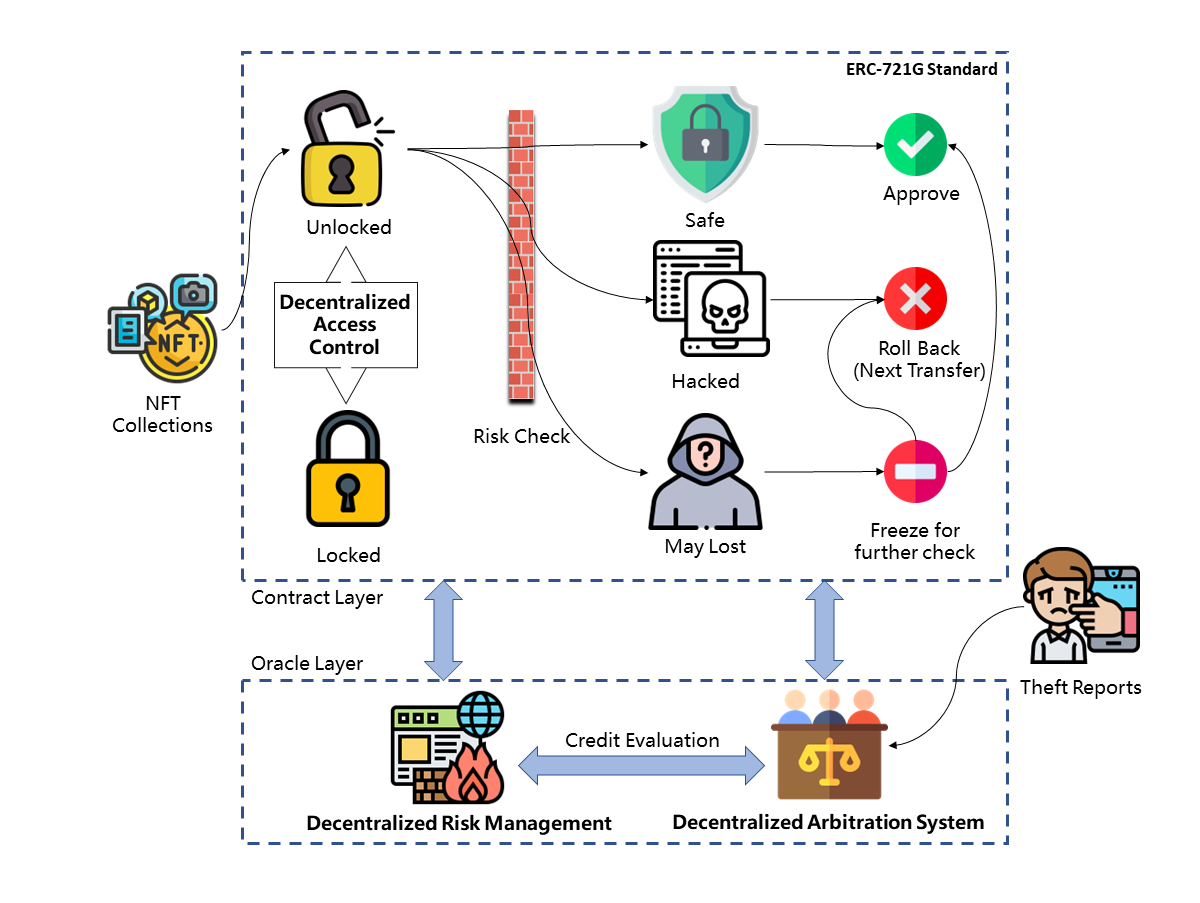}
\centering
\caption{Scheme of the TokenPatronus Mechanism} 
\label{overview}
\end{figure*}

\section{Related Work}
Even though there is a clear call for better NFT protection, this problem is generally understudied.
Existing work addresses this problem from different perspectives.

\textit{At the exchange level}. Once notified of potentially stolen tokens, cryptocurrency exchanges such as OpenSea disable the ability to buy, sell or transfer the tokens on the platform. 
However, since NFTs are stored on the underlying blockchain, they do not have the power to freeze or return stolen NFTs. 
Cryptocurrency exchanges also make use of professional account takeover protection\cite{accounttakeover} to stop trading for stolen items and automatically alarm users when an attempt to take over their account is detected or blocked.
However, the protection only takes effect on a particular platform. 
Hackers can often transfer tokens to other platforms for trading or even trade directly based on smart contracts, thus making this type of protection completely ineffective.

\textit{At the blockchain level}. Several attempts have been proposed to provide blockchain with the capability of reversing a transaction.
Fabian Vogelsteller proposed the reversible ICO\cite{reversiableico} that allows investors to take a portion of their fund back for a certain safe period. 
Dean Eigenmann proposed a reversible ERC-20 standard\cite{reversiableerc20} that allows the buyers to recall their money at any time within the 30 days escrow period. 
The \textit{Reversecoin} project is a layer-1 blockchain that introduces a timeout period between transaction initiation and confirmation. An owner can use an offline key pair to confirm or reverse a transaction within the timeout period.
The one closely related to our work is the ERC-721R standard proposed by Wang et al. \cite{erc721r}. 
Like our approach, ERC-721R relies on decentralized arbitration and extensions to ERC-721.
However, our work differs from theirs in terms of decentralized risk control, smart contract design, and the use of decentralized arbitration.

\textit{Centralized solutions}. Binance USD(BUSD) issued by \textit{Paxo}, and the USD coin issued by \textit{Circle} support a centralized capability to freeze and remove assets. 
The operator acts as a centralized judge that can reverse and nullify transactions. 
Our work is trying to provide the same functionality but in a decentralized way.

\section{Anatomy of the TokenPatronus Mechanism}

\subsection{Overview of the Architecture}
We design the TokenPatronus mechanism following the one-two-three-four principle, that is, one mechanism, two layers, three stages, and four modules. TokenPatronus mechanism covers the protection of NFT collections during the complete transaction flow, of which the scheme is shown in Fig.~\ref{overview}. When the target NFT is minted, the initial status is unlocked, and the user can choose to lock the token through the front-end page so that the contract will ban any further transactions. Once the user starts a new transaction with the NFT at the unlocked status, our DRM engine will begin to analyze the safety of this transaction. TokenPatronus will allow risk-free transfers and interrupt suspicious ones. The interrupted transactions will be frozen and reviewed by the decentralized arbitration. For the transfers that are determined to be \textit{hacked}, the risk management engine will ask the smart contract to return the NFT to previous account automatically.

Specifically, we divide the transaction of the NFT into three stages, and design the corresponding role of TokenPatronus respectively, covering pre-event protection, in-event interruption and post-event process. Four modules make up the TokenPatronus anti-theft mechanism, including the decentralized identification module, decentralized risk management engine, decentralized arbitration system and ERC-721G contract standard. To ensure the trust and the robustness, we designed four decentralization-based modules to make up the TokenPatronus mechanism, containing the decentralized access control, the decentralized risk management engine, the decentralized arbitration system and the ERC-721G smart contract standard. The details of these three stages and four modules is described in the following sections. 

\subsection{Pre-event Protection}

\begin{figure}[!htbp]
\centering
\includegraphics[width=0.5\textwidth]{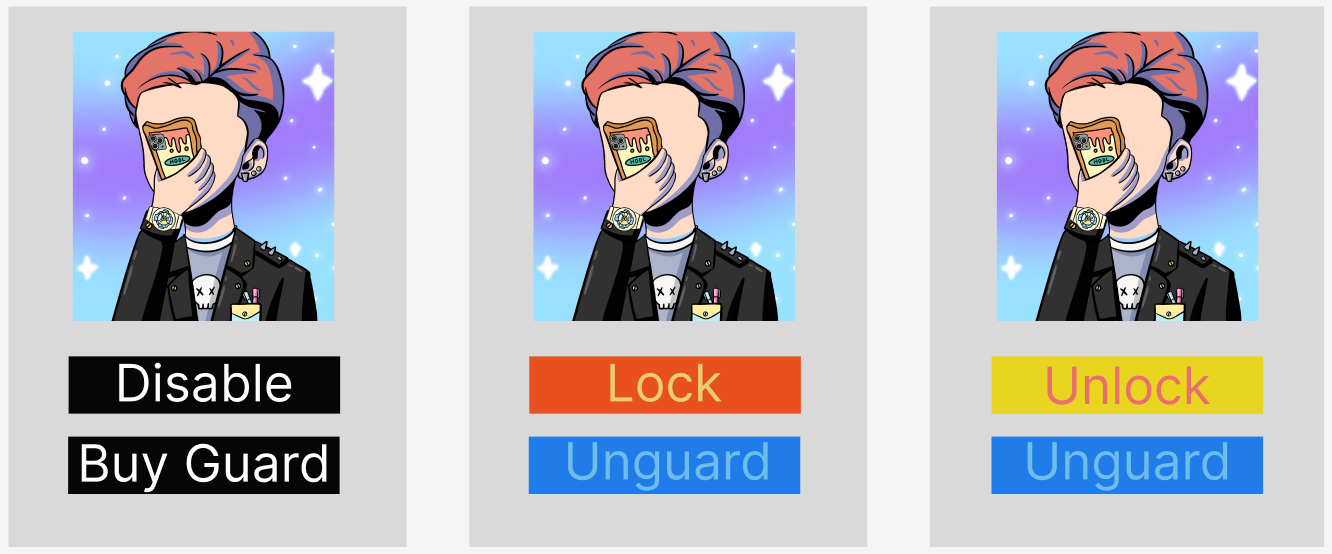}
\caption{Visual Example of the Web UI of TokenPatronus} 
\label{example}
\end{figure}

The pre-event stage represents the status when the token is  airdropped or transferred to a wallet address. The initial status of this token is unlocked, and TokenPatronus provide a function to lock it. When the owner wants to hold this NFT for a long time, he/she can choose to lock the token via the decentralized access control (DAC) module. Once the token status change to locked, any further transactions or listing on the trading market will be rejected. The verification in the DAC module requires an additional wallet signature. See Section~\ref{sec1} for more description of the DAC module. In addition, we develop a front-end web to facilitate users in managing their NFTs, with the option to easily add and unlock operations, as shown in Fig.~\ref{example}. 

\subsection{In-event Interruption}
Since NFTs can be locked in pre-protection, they cannot be sold or transferred even if a hacker has all the access to the wallet. however, there is still a chance that NFTs in an unlocked state can be lost, and we design a risk control engine at this stage to block potentially stolen NFT transactions. Specifically, each NFT transfer is passed through a DRM engine to evaluate the security of that transaction. This process is similar to the risk control system of a bank, whereas we use the oracle to implement the DRM engine. The engine outputs three status for the corresponding transaction: \textit{safe}, \textit{may lost} and \textit{hacked}. if the engine finds the transaction suspicious, it outputs the NFT state as \textit{may lost}, and the contract will freeze the NFT transaction privileges for a period of time. If the engine outputs the NFT status as \textit{hacked}, the NFT will be locked and transferred to the treasury. The real owner of this NFT has the right to recover the stolen NFTs through a decentralized arbitration system.

\subsection{Post-event Replevin}
As discussed in the previous section, we cannot guarantee that all NFTs are not stolen, nor can we guarantee that the DRM engine can output 100\% correct adjudication results. Therefore, we designed the decentralized arbitration system to handle NFTs stolen by hackers or incorrectly intercepted. Users can report a theft incident through the decentralized arbitration system, and if the system decides that the user is a victim, the NFT will be returned to the user's wallet address. In the TokenPatronus mechanism, the decentralized arbitration system, as the core part of post-event replevin, is the last anti-theft safeguard, and it has the root permission to dispose of this NFT ownership.

\section{Module Design}

\subsection{Decentralized Access Control}
\label{sec1}
\begin{figure}[!htbp]
\centering
\includegraphics[width=0.5\textwidth]{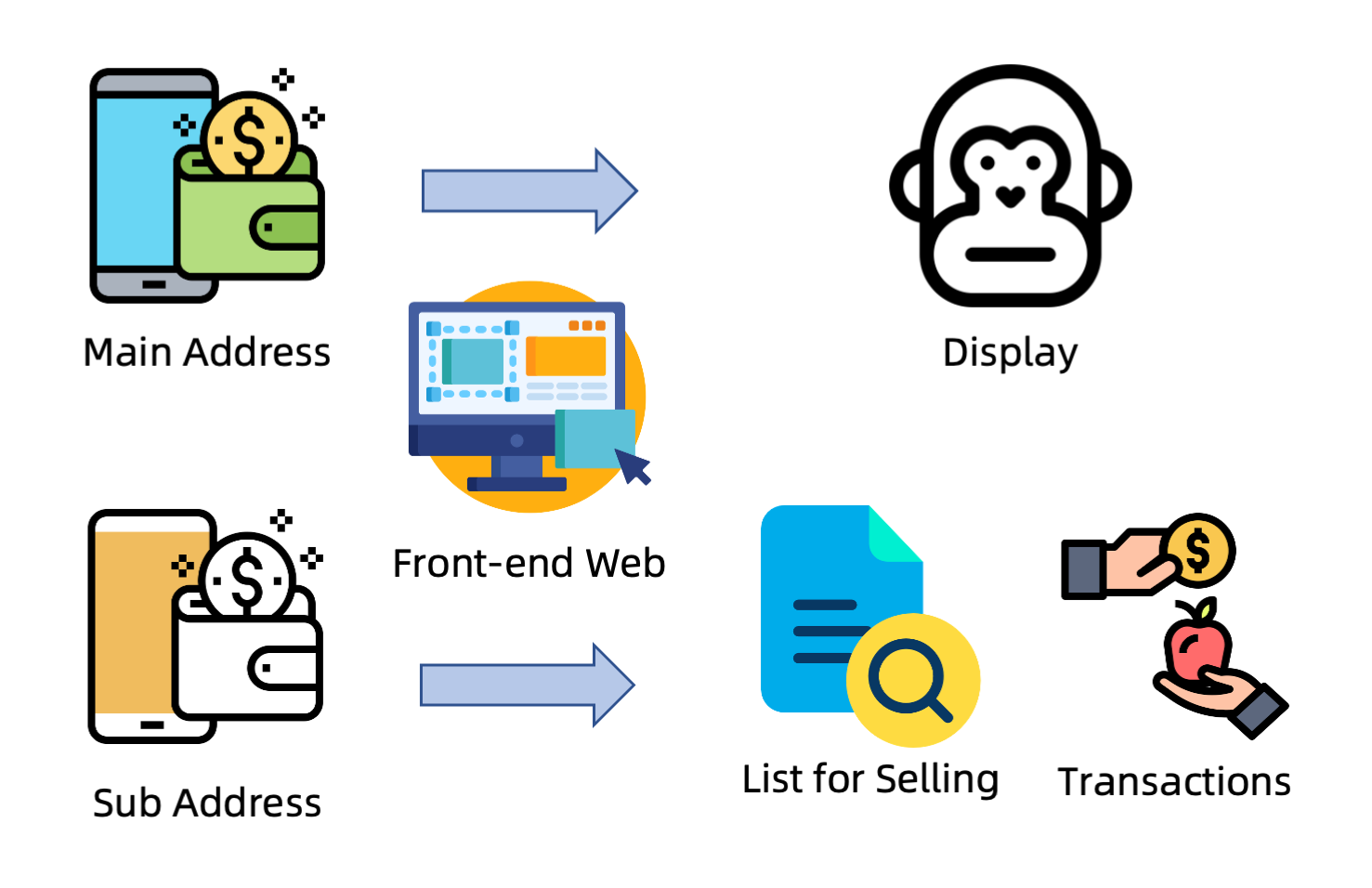}
\caption{Procedure of the Decentralized Access Control} 
\label{dac}
\end{figure}

When unlocking the NFT, the holder needs to confirm that they are operating on their own and accept the risk of transferring and transacting after unlocking, and that they bear some responsibility. If, during a transaction or transfer, the recipient does not register a sub address with the main address, then the transaction still stands. However, after the recipient receives the NFT, the NFT is locked by default and the next transfer or transaction must be registered and unlocked at the main address before it can continue. In this process, the DRM system will monitor the whole process and make risk prevention in advance. \\

As is shown in Fig.~\ref{dac}, we require a sub address to validate the ownership of the NFT. The NFT can only be displayed if the user is not registered on the front-end web. Once adding an auxiliary wallet and unlocking the NFT, the user is able to list this NFT or transfer it.

\subsection{Decentralized Risk Management}
\begin{figure}[!htbp]
\centering
\includegraphics[width=0.5\textwidth]{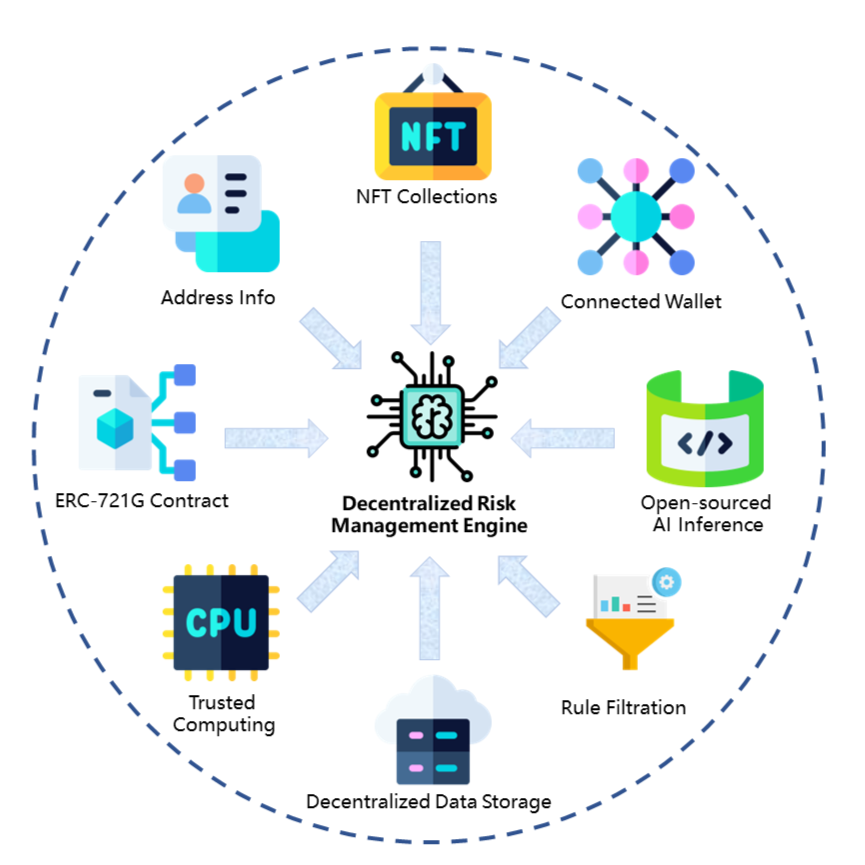}
\caption{Diagram of the Decentralized Risk Management Engine} 
\label{drm}
\end{figure}
The decentralized risk management is the key component of the TokenPatronus mechanism, which provide the necessary validation of the smart contract. The diagram of the proposed decentralized risk management engine is shown in Fig.~\ref{drm}, which is based on the the big data analysis engine. The engine evaluates the risk via eight sources, i.e., the NFT collections, address information, connected wallet information, the contract status (locked or unlocked), decentralized data storage, filtration rule, trusted computing and open-sourced AI inference. We evaluate the credit score corresponding to both sides of this transaction by obtaining the number and price of the NFT collection at the target wallet address; the contract also allows us to obtain whether there is an abnormal record of previous transactions in this NFT; and to query whether there is suspicious behaviour flagged by Etherscan\footnote{https://etherscan.io/} through the high-frequency trading accounts associated with the trading account. In addition, we have designed some rule-based filters, such as marking the trade as \textit{may lost} if the price is significantly below the floor price or the turnover rate is too high. We also design a deep learning model from the transaction history stored in the decentralized storage and open-source the corresponding inference code. The corresponding evaluation process is done via the trusted computing. The decentralized risk management will combine all the above information to assess the risk status of this transaction.

\subsection{Decentralized Arbitration System}
The Decentralized arbitration System (DAS) is a module for determining the ownership of NFTs when there is a reported loss. The victim must submit a deposit to ensure that no hacker takes advantage of the rule loophole to maliciously report the incident, and the price of the deposit fluctuates according to the value of the NFT, which is generated by a deep learning algorithm. Once the report is confirmed, the status of the corresponding NFT will be changed to \textit{may lost}. The current owner of the NFT and the person reporting the theft incident can challenge the ownership of the NFT, such as uploading conversation log as an attachment to the system. The final ownership of the NFT is decided by a vote of the members of the referee DAO. This vote follows a Byzantine fault tolerance mechanism to prevent jury members from cheating. The jury members who ultimately participate in the decision are eligible to be assigned the corresponding honor and some monetary reward is given by the system. DAS results will interact with the contract and the contract reclaims the NFT to the post-verdict wallet address and the gas fee for the transaction is borne by the reporter.

\subsection{ERC-721G Contract}

TokenPatronus interacts with two smart contracts written in Solidity. One is the ERC-721G, a ERC-721 compatible standard with anti-theft features. The other is the oracle contract, which links ERC-721G to TokenPatronus risk control engine. The oracle contract is also the only legal operator of the ERC-721G when normal users lock/unlock their NFTs and the only legal proxy of risk control engine to reclaim or judge the NFTs.\\

The design and implementation of oracle contract is trivial, with two sets of functions: \textit{request} methods and \textit{fulfill} methods. 
ERC-721G NFT holders invoke transfer related functions on ERC-721G to transfer their assets.
For a token in \textit{locked} status, every transfer operation will trigger a request from ERC-721G to the oracle contract. 
After the request is invoked, a message will be logged on chain, which will later be captured by the risk contmrol engine and activate the risk control procedure. 
Fulfill methods are only invoked if a risk is investigated to reclaim a token and freeze it for future judgement.\\

ERC-721G is not built from nothing, but based on an existed ERC-721 implementation. The difference between ERC-721G and the standard ERC-721 is that ERC-721G contains the anti-theft part, which mainly extends following section:\\

First, we add a structure to the major contract, ERC-721G:
\begin{lstlisting}
mapping(uint256 => TOKENSTATE) internal _tokenStates;
\end{lstlisting}

The \textbf{\_tokenStates} map structure holds every state of token, namely OK, LOCKED and  RECLAIMED. If \textbf{\_tokenStates[tokenID]} is OK, it means the asset can be transferred without any resistance. Otherwise, the token cannot be transferred or approved. A reclaimed token cannot be unlocked by previous holders directly, which depends on the result of judgement.

Second, we create a modifier:
\begin{lstlisting}
modifier unlockedToken(uint256 tokenId)
\end{lstlisting}
The modifier will extract the state of the token with exact tokenID, and check whether the token has a  legal state to be approved or transferred.

Third, we add the above modifier to the following ERC-721 standard methods for protecting:
\begin{lstlisting}
function approve(address to, uint256 tokenId);
function transferFrom(address from, address to, uint256 tokenId);
function safeTransferFrom(address from, address to, uint256 tokenId);
\end{lstlisting}

Finally, we also override \textbf{setApprovalForAll} method, making it supervised by risk control engine to prevent the asset from malicious (phishing)site/hacker.

\section{Conclusion}
We present TokenPatronus, an anti-theft mechanism for securing NFTs, which consists of the contract layer and the oracle layer. TokenPatronus enhances the current ERC-721 based NFT via three stages, including pre-event protection, in-event interruption and post-event replevin. In order to realize the proposed enhancement scheme, we design four modules composed of the decentralized access control, the decentralized risk management, the decentralized arbitration system, and the ERC-721G contract standard. In the future, we will continue to improve the TokenPatronus mechanism.

\section{Acknowledgments}
We thank TurtleCaseGang project owners to give insightful feedback on this paper.  

\bibliography{ref.bib}
\bibliographystyle{IEEEtran}

\end{document}